\documentclass{jaa}
\usepackage{natbib}
\usepackage{graphicx}
\usepackage{xcolor}

\def\apj{ApJ}
\def\na{NA}
\def\apjl{ApJL}
\def\aap{A\&A}
\def\mnras{MNRAS}

\def\japa{JApA}
\def\apjs{ApJ Supp}

\def\nat{NATURE}
\DeclareUnicodeCharacter{2212}{-} 
\begin{document}\sloppy

\title{LAXPC instrument onboard AstroSat: Five exciting years of new scientific results specially on X-ray Binaries}


\author{J. S. Yadav\textsuperscript{1}, P. C. Agrawal\textsuperscript{2}, Ranjeev Misra\textsuperscript{3}, Jayashree Roy\textsuperscript{3}, Mayukh Pahari\textsuperscript{4,5}  and R. K Manchanda\textsuperscript{6}}
\affilOne{\textsuperscript{1} Department of Physics, Indian Institute of Technology, Kanpur 208016, India\\}
\affilTwo{\textsuperscript{2} DAA (retd), Tata Institute of Fundamental Research, Homi Bhabha Road, Mumbai 400005, India\\}
\affilThree{\textsuperscript{3} Inter-University Centre for Astronomy \& Astrophysics, Ganeshkhind, Pune-411007, India\\}
\affilFour{\textsuperscript{4} School of Physics and Astronomy, University of Southampton, Highfield Campus, Southampton SO17 1BJ, UK\\}
\affilFive{\textsuperscript{5} Department of Physics, Indian Institute of Technology, Hyderabad 502285, India\\}
\affilSix{\textsuperscript{6} Centre for Astrophysics, University of Southern Queensland, QLD 4300, Australia\\}


\twocolumn[{

\maketitle

\corres{jsyadav@iitk.ac.in}

\msinfo{November 7, 2020}{January 6, 2021}

\begin{abstract}
With its large effective area at hard X-rays, high time resolution and having co-aligned other instruments, AstroSat/LAXPC was designed to usher in a new era in rapid variability studies and wide spectral band measurements of the X-ray binaries. Over the last five years, the instrument has successfully achieved to a significant extent these Science goals. In the coming years, it is poised to make more important discoveries. This paper highlights the primary achievements of AstroSat/LAXPC in unraveling the behavior of black hole and neutron star systems and discusses the exciting possibility of the instrument's contribution to future science.
\end{abstract}


}]


\doinum{12.3456/s78910-011-012-3}
\artcitid{\#\#\#\#}
\volnum{000}
\year{0000}
\pgrange{1--}
\setcounter{page}{1}
\lp{1}

\section{Introduction}

X-ray astronomy was born in 1962 with the chance discovery of a bright X-ray source Scorpius X-1 (Sco X-1) in a rocket flight experiment.  It is still the brightest X-ray source in the sky radiating energy in X-rays at a prodigious rate of $\sim$ $10^{36}$  ergs /sec. What process generates X-rays at such high rate, remained a mystery for many years.  In 1933 F.Zwicky and  W.Baade made a prophetic suggestion that when cores of high mass stars at the end of their life collapse, it becomes a Neutron Star in a massive explosion called supernova. The equation of state of these dense stars were investigated by J. R. Oppenheimer and G. M. Volkoff in 1939. The Neutron stars remained a theoretical curiosity till 1967 when A. Hewish and Jocelynn-Bell discovered a radio source now called  PSR B1919+21. producing periodic signal every 1.3373 sec. These pulsating objects came to be known as Pulsars, were interpreted by T. Gold as spinning Neutron stars. About a few thousand Pulsars are now known. X-rays with 33 msec pulsation period of were discovered from a source (named as PSR B0531+21) in the Crab Nebula  in 1969 in a rocket experiment. The Crab pulsar was found to be spinning down in all the wavebands monotonically. This suggests that these objects are rotation powered pulsars. 
However, these discoveries did not contribute significantly to explain the nature of Sco X-1. Soon after the launch of the first X-ray satellite UHURU in 1970, it discovered a bright X-ray source Cen X-3, whose intensity was oscillating with 4.8 sec period. The object was found to be a Binary since its light curve showed, periodic X-ray eclipses every 2.09 days. This discovery led to the conclusion that the bright X-ray sources are compact objects, either a neutron star or a black hole, in a binary with a companion star. The strong gravity of the compact objects pulls matter from the companion star. This matter forms an accretion disk from which the matter accretes to the compact object. 
                             
                             If the compact object is a neutron star, matter accretes either on the surface of the neutron star in case of the neutron star with low magnetic field ($\leqslant$ 10$^{9}$ gauss),  or accretes at the magnetic poles   of a highly magnetized neutron star. If the magnetic axis is offset from the rotation axis, the X-ray intensity oscillates as the neutron star spins. Several hundred X-ray binaries have been discovered in the Galaxy, majority of them with neutron star as the X-ray source. The in falling matter at the poles under strong gravity of the neutron star gets heated up as it settles down on the surface of the neutron star or at its magnetic poles. The X-ray luminosity of the binary is due to release of gravitational energy and depends on the accretion rate.

\subsection{Classification of X-ray Binaries and their Salient Features :}
                            The X-ray binaries are classified in two categories; the Low Mass X-ray Binaries (LMXBs) and the High Mass X-ray Binaries (HMXBs). The LMXBs usually have a low mass ($\le$ 2 Solar mass) companion star while the HMXBs usually have a massive companion star ($\geqslant$ 5-30 Solar Mass). 
 In the X-ray binaries, the compact objects may be either a neutron star (NSXBs) or a black hole (BHXBs).
  Among all the wavebands, the most rapid and violent brightness changes are observed in the X-ray emitting sources. X-ray binaries  (NSXBs or BHXBs) exhibit intensity variations over a wide range of time scales ranging from millisecond to months and years. 

 In a majority of the neutron star LMXBs, the magnetic field is weak ( $\leqslant$ $10^{9}$ gauss) but a small number of the neutron star LMXBs  have higher magnetic fields $\geqslant$ $10^{9}$ Gauss ($\sim$ $10^{12}$ Gauss). The well known binary Her X-1 is an example of the LMXBs in which the neutron star has a high magnetic field as inferred from the presence of the cyclotron feature in its energy spectrum.  These systems also exhibit regular X-ray pulsations similar to those in the HMXBs. The LMXBs 
with low magnetic field neutron stars, are thought to have evolved from high magnetic field neutron star binaries in which the accretion has spun up the neutron stars over a long period during which the magnetic field has significantly decayed. The accreting millisec pulsars (spin periods $\sim$ msecs) are an example of the spun up neutron stars from their progenitor LMXBs.  The LMXBs  with weak magnetic field, are characterised by short orbital periods and  sporadic or regular short X-ray bursts (strictly not periodic), that have a typical duration of about a few seconds to a  minute or in rare cases longer.

 The matter accretes from the companion star on the neutron star via Roche Lobe overflow leading to the formation of an accretion disk around the neutron star.  From the inner part of the accretion disk, the matter falls on the surface of the neutron star. The accreted matter piles up on the surface of the neutron star forming a layer with increasing thickness. As the accretion rate is variable, thickness of the layer in not uniform. When the accumulating matter on the neutron star surface develops density and temperature conditions in a localized region that ignite thermonuclear reactions a powerful X-ray Burst occurs. These are called thermonuclear bursts. Initially the flame is localized in a spot but it spreads quickly over the neutron star surface. Due to uneven spread of the nuclear burning, the X-ray in-tensity is modulated due to rapid spin of the neutron star. 
The intensity oscillations with frequency of several hundred to about a kHz are detected in the LMXB bursts. This is one of the ways to measure the spin periods of the neutron stars in the LMXBs.
                            
Using the RXTE/PCA data,  kHz QPOs  were discovered from about a dozen LMXBs (usually occurring in pairs) \citep{2001AIPC..599..406V}. The precise origin of kHz QPOS is still not well understood. The LMXBs have multi-component continuum energy spectra consisting of a disk blackbody component and a Thermal Compton power-law.  The thermal Compton part of the spectrum is most likely due to the Compton up scattering of X-rays by hot electrons in a halo of accreted material surrounding the disk.

A precise modeling of the continuum is essential to detect weak absorption features known as Cyclotron Resonant Scattering Features (CRSFs) or commonly referred as Cyclotron lines, that originate in the accretion column above the magnetic poles. The CRSFs are detected in the strong magnetic field HMXBs as well as in the spectra of the strong field LMXBs.  In the neutron stars with strong magnetic fields, the accreting matter is guided from the disk by the magnetic field lines of the neutron star to its magnetic poles. The X-ray spectrum emitted by the plasma in the accretion column is affected by the magnetic field. This interaction gives rise to the CRSFs. The energy of Cyclotron lines provides a direct measure of the dipole magnetic field of the neutron star in the accretion column region. 
 
                      The HMXBs usually have a pulsating X-ray source. The pulsation periods have a range from $\sim$ 1 sec to  hours. The slowest pulsar 2S 0114+65 has a spin period of 2.7 hours. The spin rate of the pulsars is variable and depends on the accretion rate. A study of the evolution of the pulsation periods of many pulsars, shows that most pulsars exhibit both spin up as well as spin down episodes.  Depending on the accretion rate, a pulsar may make transition from spin up to spin down and vice versa. In a class of X-ray Binaries, the optical star is a Be star with a shell or disk of matter ejected by the companion. Whenever the neutron star crosses the shell or disk during the course of its orbital motion, the accretion rate shoots up resulting in dramatic increase in brightness known as X-ray outbursts that occur once or twice in an orbital period. Some times there is a sudden catastrophic release of matter from the disk or shell of the Be star. This may be due to thermal instability that has still not been well understood. This leads to a massive spike in the accretion rate resulting in a giant outburst which is non-periodic and unpredictable. This phenomenon also manifests in many X-ray transients that remain in a dormant state with an exceedingly low accretion rate rendering them invisible. This hibernation may last for years and even decades. Then due to reasons still not understood, they spring back to life with giant X-ray outbursts lasting for tens of days to months. QPOs of  $\sim$ mHz to tens of Hz have been detected in many  HMXBs. Study of these QPOs provides insight into the radiation environment closest to the neutron star. 
                                                      
                            In black hole X-ray binaries (BHXBs), accretion disk and relativistic radio jets are integral part  on all black hole mass scales and  they provide simple scaling of time and length with  the mass of the black hole from supermassive black holes in active galactic nuclei (AGNs)  to stellar mass black holes in  BHXBs  in our Galaxy. When accretion rate from the companion star is not sufficient to support continuous accretion flow to the black hole, matter fills the outer disk  until a critical surface density is reached and an outburst is triggered. An outburst in the BHXBs may last from  $\sim$ 20 days to several months.  There are over 20 confirmed BHXBs and many more candidates \citep{remillard2006x}. Out of 20 BHXBs, 17 are transient X-ray sources (mostly LMXBs) and the rest three are persistent bright X-ray sources which have a  massive  O/B type  stars as their companion (HMXBs) \citep{remillard2006x}. Many new BHXBs have been discovered with time \citep{2020ApJ...889L..17M, 2019MNRAS.488..720B}, and  MAXI shows now lightcurves of around 40 BHXBs. 

The BHXBs show different X-ray states at different stages of an outburst. Initially, luminosity was the sole  criteria to determine X-ray states \citep{van1994book, tanaka1995x}.  
With the availability of larger data sample during RXTE/PCA era, it became clear 
that X-ray states are not simple function of the luminosity  \citep{lewin2006compact, remillard2006x}. 
It is suggested that beside  the mass accretion rate,  there must be at least one additional 
parameter which drives X-ray state transition. The hardness-intensity diagram (HID) which shows a 
q-shaped track, has been used to study the evolution of outbursts and associated 
X-ray states in the BHXBs systems.  The four distinct  X-ray states identified in the HID are 1. Low Hard state (LS), 2. High Soft state (HS),  3. Hard Intermediate State (HIMS), and 4. Soft Intermediate state (SIMS) 
\citep{belloni2006black}. The LS state is associated with relatively low accretion rate (lower than other bright  X-rays states) but can be observed at all luminosity. It is a radio loud state. The energy spectrum is dominated by a hard power-law while power  spectrum shows a strong band-limited noise. The HS state is strongly dominated by thermal disk component. Radio quenching is seen during this state. The intermediate states (HIMS $\&$ SIMS) have significant contributions from both the hard power law and the thermal disk components and show the most complex variability characteristics including most of the quasi-periodic oscillations(QPOs). Transient radio jets are seen during 
HIMS to SIMS transition.  \citet{remillard2006x} provided alternative scheme of X-rays states in BHXBs in terms of  Hard State (HS), Thermal State (TS) and Steep Power Law (SPL) state. Here major difference is the SPL state which has significant contributions from Thermal as well as non-thermal emissions.

\subsection{LAXPC Instrument and its features to study Temporal and Spectral properties :}

\begin{figure}
\begin{center}
{\includegraphics[width=1.0 \linewidth,angle=0]{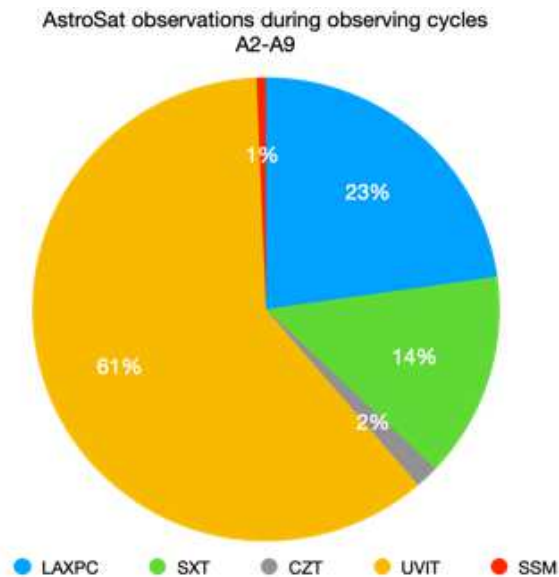}}
\end{center}
\caption{Astrosat observations during observing cycles A2-A9 showing percentage of time devoted to  different instruments as primary instrument.}
\label{fig1}
\end{figure}

            Large Area X-ray Proportional counter (LAXPC) instrument is one of the major instruments onboard Astrosat which is India`s first space science  observatory  \citep{2006AdSpR..38.2989A}. There are three X-ray instruments on-board AstroSat which cover a wide energy band. These X-ray instruments are: (i)  LAXPC instrument \citep{2016SPIE.9905E..1DY, 2017JApA...38...30A, 2017CSci..113..591Y} covering 3-80 keV region, (ii) a Cadmium-Zinc-Telluride Imager (CZTI; \citet{2017JApA...38...31B}) array covering  30-100 keV, and  (iii) a Soft X-ray Imaging Telescope (SXT; \citet{2016SPIE.9905E..1ES}) covering  0.3 - 8 keV. Beside these X-ray instruments, there is an Ultra-Violet Imaging telescope (UVIT: \citet{2017JApA...38...28T}). All the above instruments are co-aligned to provide simultaneous  multi wavelength observations from optical to hard X-ray. There is also a  Scanning Sky Monitor (SSM) onboard Astrosat.
The LAXPC instrument  uses three co-aligned identical LAXPC detectors (LAXPC10, LAXPC20 and LAXPC30) to achieve  large area. The performance of LAXPC detectors are discussed by \citet{2020JAPA.tmp.antia} in the same issue. A high detection efficiency in the entire 3-80 keV energy region is achieved by having a 15 cm deep detection volume divided in 5 identical Anode Layers and filling the Xe+CH$_{4}$ gas at two atmosphere pressure (1520 torr).  
 
          The LAXPC instrument was designed to meet the objectives of studying the intensity variations with time, have a high detection efficiency in 3-80 keV band, provide a large effective area over the broad energy range  to be able to study weak ($\sim$ a few milliCrab) sources and have a good spectral resolution to measure the X-ray continuum spectra of X-ray binaries with precision for deciphering the presence of weak cyclotron lines in the NSXBs. A unique feature of the LAXPC is that every detected photon is time tagged to an accuracy of 10 $\mu$sec to enable investigation of rapid intensity variations over even sub-millisec scale. This feature provides LAXPC with the capability to measure high frequency QPOs of even a few kHz.  Good energy resolution (FWHM $\sim$ 15\% at 60 keV) was achieved by using an onboard gas purifier system and continuously monitoring of the resolution of the Veto layer by shining 60 keV X-rays on it from an Am$^{241}$ source. We have done extensive lab as well as in orbit calibration of LAXPC detectors \citep{2017ApJS..231...10A}.
 Further improvements and updates on LAXPC detector calibration and background are done \citep{2020JAPA.tmp.antia, 2020JAPA.tmp.misra} (this issue).  
                                
           Since October 2016, AstroSat observatory has operated on the basis of targets selection through scientific proposals.  during the observing cycles A02-A09 in the last 5 years. A comparison chart for the observations with different instruments as primary instrument is shown in Figure \ref{fig1}. Primary instrument is defined in the Astrosat  science proposals (as science proposal requirement).  It may be noted that  simultaneous data from all the instruments will be available for a given source pointing if all the instruments are on during the observation.

\section{Highlight of Results on Neutron Star X-ray  Binaries (NSXBs):}
Large number of X-ray binaries with an accreting Neutron star as the X-ray source, have been observed with the LAXPC instrument to investigate their timing and spectral characteristics.  Due to various reasons, briefly discussed in Antia et al( 2020),  less than 50\% data of the sources have been analyzed so far. Nevertheless, several new and interesting results have emerged from these studies. Here we summarize highlights of some of these results.

\begin{figure}[h]
\begin{center}
{\includegraphics[width=1.0 \linewidth,angle=0]{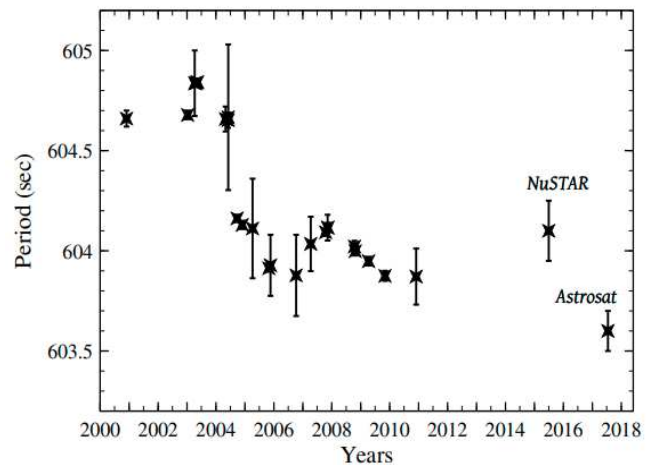}}
\end{center}
\caption{Evolution of the Pulsar period of 4U 1909+07 during 2001-17.}
\label{fig2}
\end{figure}

\subsection{Timing Properties : Spin Periods and their evolution, Low and High Frequency QPOs.}

 The spin periods of the neutron stars are affected by the accretion rate and the resulting accretion torque that varies with times. Usually the low and high frequency QPOs are found in the LMXB pulsars while the low frequency QPOs are more common in the HMXB pulsars. Astrosat has studied several neutron star binaries to investigate their spin rates and their evolution with time. LAXPC has observed a large number of HMXBs to investigate their periodic and aperiodic variations like determination of the spin periods and their time evolution, QPOs and their origin We summarize some of the significant results derived from Astrosat. 

The LMXB  3A 1822-371 has an orbital period of 5.2 hours and pulsates with 0.59 sec period \citep{2001ApJ...553L..43J}. The spin period from 1996 and 1998 data leads to a spin-up rate of -2.85$\times$ $^{-12}$ s s$^{-1}$. Subsequently measurements of the spin period from several satellite observations have shown that the pulsar is monotonically spinning up implying that the accretion rate has been constant over $\sim$ 20 years. Astrosat LAXPC observations of 3A 1822-371 in September 2016 yielded a barycenter corrected pulsar spin period to be Pspin =0.5914907$\pm$0.0000003 s confirming that the pulsar is still spinning up \citep{2020JAPA.tmp.amin} Making a linear fit to all the measured spin periods, gives a spin up rate=(-2.62$\pm$0.02)$\times$10$^{-12}$ s s$^{-1}$ giving a spin up timescale (P/P)=7169 yr in agreement with the previous results reported by \citet{2010MNRAS.409..755J}.

Astrosat also studied accretion powered 2.26 millisecond pulsar SAX J1748.9-2021, when a short and faint outburst occurred in it in 2017, with the SXT and LAXPC instruments. From the spectral and timing analysis of the LAXPC data,,the best-fitting orbital solution for the 2017 outburst was derived. Using this an average local spin frequency of 442.361098(3) Hz was obtained. The pulse profile obtained in 3-7 keV and 7-20 keV gave a constant fractional amplitude $\sim$ 0.5\% in contrast to earlier reported energy dependent profile. The combined SXT+LAXPC spectrum in 1-50 keV showed the source to be in a hard spectral state. This spectrum is best fitted with a single temperature blackbody and thermal Comptonization model. Time resolved analysis of the bursts revealed complex evolution in emission radius of blackbody for the second burst suggestive of a mild expansion of photospheric radius.

The HMXB Pulsar 4U 1909+07 studied with the Astrosat-LAXPC observation of 2017 July, showed presence of  604 sec X-ray pulsations. Using the spin period measurements obtained earlier with the various X-ray satellites and the period deduced from the present work, a plot of Spin Period versus Time is shown in Fig \ref{fig2} \citep{2020MNRAS.498.4830J}. Despite a brief episodes of spin-down,  there is a clear long term trend of spin-up. The pulse period changed from 604.7 s to 603.6 s between 2001-2017 resulting in an average spin-up rate of 1.71$\pm$10$^{-9}$ ss$^{-1}$. The pulse profiles show strong energy dependence evolving from a complex broad structure in soft X-rays into a profile with a narrow emission peak followed by a plateau in energy ranges above 20 keV. The change in the pulse profile suggests a change in the beaming pattern. 

\begin{figure}[b!]
\begin{center}
{\includegraphics[width=1.0 \linewidth,angle=0]{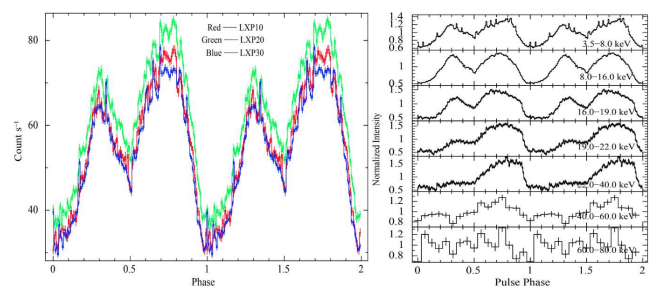}}
\end{center}
\caption{The pulse profile of 4U 1907+09 has double peak shape shown in the left panel. The energy resolved pulse profiles created from combined data of the three LAXPCs are shown in the right panel.} 
\label{fig3}
\end{figure}

Another HMXB 4U 1907+09 with a spin period of $\sim$ 404 sec was studied by by using the LAXPC observations of 4th and 5th June (2020). Timing analysis of the LAXPC data yielded a period of 442.33$\pm$0.07 sec. Two peaks of the pulse profile of 4U 1907+09 shown in Fig \ref{fig3} (left panel), exhibited different energy dependence, the pulsed fraction of the main peak increased till about 40 keV and decreased after that while the secondary peak disappeared at energy above about 20 keV. Energy resolved pulse profiles created from combined data of the three LAXPCs are shown in Fig \ref{fig3} (right panel). The pulsar spin-down was found to continue \citep{2019ApJ...880...61V}.

\begin{figure}
\begin{center}
{\includegraphics[width=0.6 \linewidth,angle=-90]{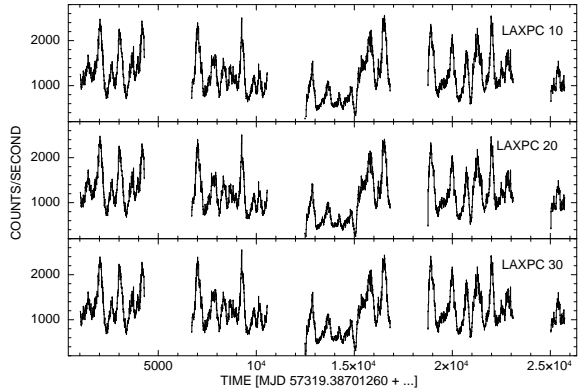}}
\end{center}
\caption{The background subtracted light curves from the 3 LAXPCs showing clear intensity oscillations of $\sim$ 1 mHz and $\sim$ 1.7 mHz in the X-rays from the HMXB Pulsar 4U 0115+63}
\label{fig4}
\end{figure}

The Be binary with 194 sec pulsar GRO 12058+42 underwent a massive outburst in April 2018 and was observed by Astrosat to investigate its timing and spectral features. The SXT and LAXPC showed strong pulsations with a period of 194.2201$\pm$0.0016 s, and a spin-up rate of (1.65$\pm$0.06)$\times$10$^{-11}$ Hz s$^{-1}$ \citet{2020ApJ...897...73M}. Pulse profiles in 3-80 keV were found to be energy dependent. The Power Density Spectrum (PDS) of the source revealed a 0.090 Hz QPO and its higher harmonics.  Another HMXB pulsar OAO 1657-415 with 10.4 day orbital period, was studied by \citet{2020JAPA.tmp.jaisawal} with two LAXPC observations in March and July 2019. The observations covered orbital phases of 0.681-0.818 and 0.808-0.968. Despite being outside the eclipsing regime, the PDS from the first data did not show any clear signature of pulsation or quasi-periodic oscillations. However, during the second observation, the X-ray pulsations at a period of 37.0375 s. were clearly detected in the orbital phase range 0.808-0.92. The pulse profile of the pulsar from the second observation consisted of a broad single peak with a dip-like structure in the middle all across the observed energy range. The evolution of the emission geometry was probed by constructing the pulse profile in narrow time and energy segments. The energy spectrum of OAO 1657-415 is approximated by an absorbed power-law model with an iron fluorescent emission line. These results are explained in the frame of stellar wind accretion model.

Another HMXB X-ray binary pulsar 3A 0726-260 (4U 0728-25) was observed on 2016 May 6-7 with the LAXPC and SXT, after a gap of almost 20 years. The light curves of the binary show Strong X-ray pulsations with a period of 103.144$\pm$0.001 seconds in 0.3-7 keV with the SXT and in 3-40 keV with the LAXPC. The pulse profiles are energy dependent, and there is an indication that the pulse shape changes from a broad single pulse to a double pulse at higher energy. At energies above 20 keV pulsations with a period 103.145$\pm$0.001 seconds are detected for the first time and a double peaked pulse profile is observed from the source. The energy spectrum of the source is derived from the combined analysis of the SXT and LAXPC spectral data in 0.4-20 keV. The best spectral fit is obtained by a power law model with a photon index (1.7$\pm$0.03) with high energy spectral cut-off at 12.9 $\pm$ 0.7 keV and a broad Iron line at $\sim$ 6.3 keV \citep{2020RAA....20..155R}. 
 
Detection of $\sim$ 1 mHz and $\sim$ 1.7 QPO in the Be X-ray Binary 4U 0115+ 63 were reported by \citet{2019ApJ...872...33R} using the Astrosat-LAXPC observation of the HMXB. The QPOs were detected on 2015 October 24 during the peak of a giant type II outburst. Prominent intensity oscillations seen at $\sim$ 1 and$\sim$ 1.7 mHz are shown in Fig \ref{fig4}. Cyclotron resonant absorption feature at 11 keV and its 3 harmonics are detected in the energy spectra. Possible models to explain the origin of the $\sim$ mHz oscillations are examined. These oscillations bear resemblance to the intensity oscillations observed from some other neutron star and black hole sources and may have a common origin. Current models to explain the instability in the inner accretion disk causing the intense oscillations were examined and found to be inadequate.

\subsection{Energy Spectra, Detection and Study of CRSFs:}

LAXPC has observed a large number of HMXBs to investigate their periodic and aperiodic variations and measure their continuum spectra and their characterization to reveal the presence of Cyclotron lines (CRSFs). Thus far Astrosat-LAXPC has discovered Cyclotron lines in a few pulsars in which earlier there was no report of the presence of CRSF. Study of cyclotron line energy and its profile enable determination of the magnetic field of the neutron star and at the site of their origin in the accretion columns from the following relation

\begin{equation}
E_{c}=11.6(B/10^{12} G)(1+z)^{-1} keV
\end{equation}                                               

E = Energy of the Cyclotron line, B = Magnetic field of the neutron star in unit of 10$^{12}$ gauss and z is gravitational red shift due to neutron star.

There are reports of shift of the line energy with time and correlation of this with the X-ray luminosity. The results on this are conflicting and reality of variation of line energy is still an open question.

\begin{figure}
\begin{center}
{\includegraphics[width=1.0 \linewidth,angle=0]{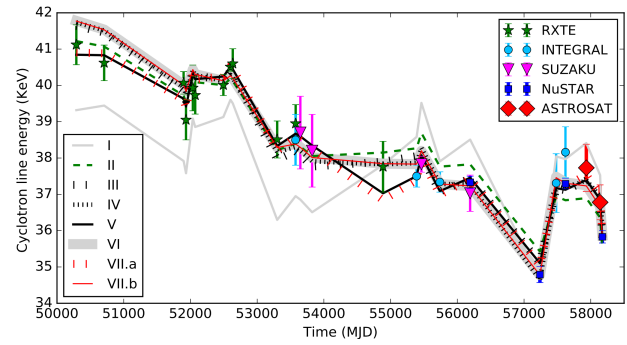}}
\end{center}
\caption{Shift in the energy of the Cyclotron line in Her X-1. The 2018 observation with the LAXPC gave a value of 37.5 keV.}
\label{fig5}
\end{figure}

Her X-1 was the first source in which \citet{1978ApJ...219L.105T} discovered a CRSF usually termed as Cyclotron line at $\sim$ 40 keV. The CRSF energy was found to vary with pulse phase, X-ray luminosity, the phase of  35-day precession cycle and with time \citep{2014A&A...572A.119S}. Using data acquired from several X-ray satellites over a 20 year period, the CRSF energy was inferred to decline by 4.5 keV from 1996 till 2012 \citep{2014A&A...572A.119S} as seen in Fig \ref{fig5} from \citet{2020MNRAS.497.1029B}. After 2015 the trend was reversed and the CRSF energy began to recover reaching a value of 37.4 keV \citep{2019A&A...622A..61S}. Recent analysis of Her X-1 observations in 2018 with the LAXPC shows that since then the CRSF energy appears to be constant around 37.5 keV \citep{2020MNRAS.497.1029B}.  

\citet{2020JAPA.tmp.amin} constructed the energy spectrum  of the pulsar 3A 1822-371 using LAXPC10 and LAXPC20 data and fitted it using 3 different continuum models. In all the cases an absorption feature at 23 keV was present which is suggested to be a CRSF and if correct this implies a magnetic field strength of B = (2.7-3.4) $\times$ $10^{12}$ Gauss for the neutron star.   
The energy spectrum of HMXB 4U 1907+09  (spin period of $\sim$ 404 sec) was studied by by \citep{2019ApJ...880...61V} using the LAXPC observations of 4th and 5th June (2020). A CRSF was detected  in the spectrum at 18.5$\pm$0.2 keV . Pulse phase-resolved spectroscopy was carried out with 10 independent phase bins and variations of the CRSF parameters with pulse phase were found to be consistent with previous studies.

The pulsar GRO J2058+42 was observed with the LAXPC. Its energy spectrum extracted from the LAXPC data revealed presence of 3 CRSFs \citep{2020ApJ...897...73M}. The cyclotron absorption features were detected in (9.7-14.4) keV, (19.3-23.8) keV and (37.8-43.1) keV, one of which is the fundamental  line and the other 2 are harmonics The pulse phase resolved spectroscopy of the source showed phase dependent variation in line energy and relative strength of these features.       
An important finding from Astrosat is the discovery of a Cyclotron line at $\sim$22 keV in the LAXPC energy spectrum  by of the HMXB pulsar 4U 1538-52 which has an orbital period of 3.75 days and  the spin period is currently $\sim$ 527 sec \citet{2019MNRAS.484L...1V}.The Pulse profile is double peaked at low energy and has a single peak in high energy range, the transition taking place around the cyclotron line energy of the source.

\begin{figure}
\begin{center}
{\includegraphics[width=1.0 \linewidth,angle=0]{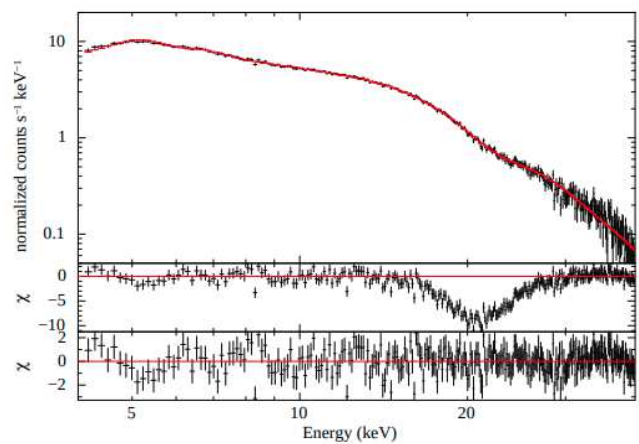}}
\end{center}
\caption{The energy spectrum of the pulsar 4U 1538-52 derived from the LAXPC data. Top panel shows the observed spectrum by the Black points and the best fitted spectrum by a Red line. The middle panel shows Residuals when one fits only the continuum spectrum and the bottom panel shows fit with continuum and a Cyclotron line at 22 keV.
}
\label{fig6}
\end{figure}

The CRSF is detected with a very high significance in the phase averaged spectrum shown in Fig \ref{fig6}. It is one of the highest signal to noise ratio detection of CRSF for this source. A detailed pulse phase resolved spectral analysis with 10 independent phase bins was performed and  parameters of the continuum spectrum and CRSF parameters were derived. Theses show pulse phase dependence over the entire phase with a CRSF energy variation of $\sim$13\% in agreement with previous studies. An increase in the centroid energy of the CRSF observed between the 1996-2004 (RXTE) and the 2012 (Suzaku) observations, is confirmed affirming that the increase in the line energy was a long-term change.

AstroSat observed the Be/X-ray binary pulsar SXP 15.3 in the Small Magellanic Cloud (SMC) during its outburst in late 2017, when the source reached a luminosity level of $\sim$ $10^{38}$ erg s$^{-1}$. Timing and spectral analysis between 3 and 80 keV lead to the pulse profile that exhibits a significant energy dependence. The pulse shape changes from a double peaked profile to a single broad pulse at energies $>$ 15 keV. The energy spectrum suggests presence of a Cyclotron Resonance Scattering Feature (CRSF) at $\sim$5 keV and is independent of the choice of the continuum model. This feature is also detected in the spectrum obtained by the NuStar. Till now CRSF has been reported in about 36 X-ray pulsars in binaries \citep{2017JApA...38...50M, 2019A&A...622A..61S}. To the best of our knowledge this is the lowest energy Cyclotron line confirmed in any pulsar. This indicates a magnetic field strength of 6$\times$ $10^{11}$ G for the neutron star \citep{2018MNRAS.480L.136M}.

\subsection{Thermonuclear Bursts and QPOs in LMXBs:}
The thermonuclear bursts (or commonly called as Type-1 bursts) occur in the LMXBs having a weakly magnetized ($<$ 10$^{9}$ G) neutron star. Detailed timing and spectral studies of these bursts provides information about the spin period, temperature of the thermonuclear burning surface and radius of the neutron star. The event mode data in which each photon is time tagged to an accuracy of about 10 microsec, enables studies of high frequency phenomenon like kHz oscillations, Some significant results on the Type-1 bursts observed with the LAXPC are summarized below. 

LAXPC capability for high time resolution studies of phenomenon like detection of Coherent Oscillations , kHz QPOs etc has been demonstrated by \citet{2017ApJ...841...41V}. They analyzed the LAXPC observations of 8th March 2016 for the LMXB source  4U 1728-24. A 3 ks data stretch revealed occurrence of a typical Type-1 burst of about 20 sec duration. Dynamical power spectrum of the data in the 3-20 keV band, shows presence of Coherent burst oscillation whose frequency increased from  361.5 to 363.5Hz consistent with an earlier result showing the same spin frequency of the neutron star. Our knowledge of the spin periods of the neutron stars in LMXBs can also be derived from detection of the Coherent Burst Oscillations in the initial phase of the bursts. A kHz QPO, whose frequency drifted from $\sim$ 815 Hz  to $\sim$ 850 Hz, was detected just before the burst. The QPO was detected below 10 keV as well as in 10-20 keV band, which was not possible with the RXTE. Even for a short observation with a drifting QPO frequency, the time-lag between the 5-10 and 10-20 keV bands can be constrained to be less than 100 microseconds.     

\citet{2019MNRAS.482.4397B} studied the LMXB 4U 1636-536 using 65 ks LAXPC observation over 2 days and detected seven thermonuclear X-ray bursts including a rare triplet of X-ray bursts. Time resolved spectroscopy performed during these seven X-ray bursts suggested the presence of Photospheric Radius Expansion in three of these X-ray bursts. A transient QPO at 5 Hz was also detected. No evidence of kilo-hertz QPOs or coherent burst oscillations, was found in the bursts and may perhaps be due to the hard spectral state of the source.

Effects of Thermonuclear X-ray burst on non-burst emissions in the soft state of the LMXB 4U 1728-34 was investigated by \citet{2018ApJ...860...88B} to understand if a significant fraction of the burst emission, which is reprocessed, contributes to the changes in the persistent emission during the burst. This is important since it can introduce significant systematics in the neutron star radius measurement using burst spectra. Analyzing the bursts data for 4U 1728-34 in the soft state it was concluded that the burst emission is not significantly reprocessed by a corona covering the neutron star.

\citet{2021NewA...8301479D} detected 5 Type-1 thermonuclear X-ray bursts and one burst-like event in the neutron star LMXB source Cygnus X-2 using LAXPC X-ray data. An energy resolved burst profile analysis and time resolved spectral analysis for each of the bursts was performed to characterize the burst properties. An evolution of the blackbody temperature and radius is also observed during each burst. A search for Coherent Burst Oscillations gave only upper limits. A study of the hardness-intensity and color-color diagrams show that during the 2016 LAXPC observation, Cygnus X-2 was in the early Flaring Branch (FB).

The LMXB 4U 1323-62 is an interesting source from which periodic X-ray dips were first detected by EXOSAT \citet{1985Natur.313..768V, 1989ApJ...338.1024P}. It was extensively studied by PCA/ RXTE and a study of 40 Type 1 bursts showed a recurrence time of 2.45-2.59 hr. which is probably the orbital period of the binary. The LMXB 4U 1323-62 was observed with the LAXPC and detected 6 Type-1 thermonuclear X-ray bursts in $\sim$ 50 ks exposure. The time gap between  the successive   thermonuclear bursts was found to be consistent with the orbital period. Using the linear and  orbital quadratic ephemerides, the value of the Orbital period is estimated to be  2.65 to 2.70 h.   
The gap observed between the bursts in two case, is nearly double the wait time of consecutive bursts as 2 bursts were missed in the data gaps. The nearly  fixed time gap is the time required to accumulate the accreted matter to reach the level at which the thermonuclear reactions set in producing a burst. The light curve of 4U 1323-62 also revealed the presence of two dips. A known low Frequency QPO (LFQPO) was detected at $\sim$1 Hz, from the source. However, no evidence of kHz QPO was found. The radius of the blackbody is found to be consistent with the blackbody temperature and the black-body flux of the bursts \citep{2020RAA....20...98B}.

The spectral and timing properties of the atoll source 4U 1705-44 were studied by \citet{2018MNRAS.477.5437A} using 100 ks data from the LAXPC instrument. The source was in the high-soft state during the LAXPC observations and traced out a banana track in the Hardness Intensity Diagram (HID). From the Power Density Spectra (PDS) a broad Lorentzian feature centered at 1-13 Hz and a very low frequency noise (VLFN) is detected. The energy spectra are well described by sum of a thermal Comptonized component, a power-law and a broad (FWHM $\sim$2 keV) Iron line having equivalent width (EW $\sim$ 369-512 eV). Only relativistic smearing in the accretion disc can not explain the origin of this feature. A systematic evolution of the spectral parameters eg optical depth, electron temperature etc. is seen as the source moves along the HID. Search of correlation between frequency of the Broad Lorentzian and the spectral parameters seems to suggest that the frequency varies with the strength of the corona.

The  structure of Corona of the well known  Z source GX 17+2 was studied by \citet{2020MNRAS.tmp.2756M} in 0.8-50 keV using the Soft X-ray Telescope (SXT) and the  LAXPC data. For the first time, Cross-correlation studies were performed using SXT soft and LAXPC hard light curves and they exhibited correlated and anti-correlated lags of the order of a hundred seconds. Spectral modeling gave  disk radius of $\sim$ 12-16 km indicating that disk is close to the ISCO and a similar value of disk radius was deduced based on the reflection model. Corona size was inferred to be 27-46 km and 138-231 km depending on the model used.  Size of the X-ray emitting  Boundary Layer (BL) was determined to be 57-71 km. The observed lags and no movement of the inner disk front strongly indicates that the varying corona structure is causing the X-ray variation in the Normal Branch (NB) of Z source GX 17+2. 

\citet{2020RAA....20...98B}  investigated evolution of  timing and spectral properties of  the bright Z-source GX 5-1  using  February, 2017 observations with the SXT and LAXPC instruments. The 0.8-20 keV spectra from simultaneous SXT and LAXPC data at different locations of the hardness-intensity plot is well described by a disk emission and a thermal Comptonized component. The  disk flux ratio (ratio of the disk flux to the total flux) increases monotonically along the horizontal branch to the normal one. Thus in the normal branch, the disk dominates the flux while in the horizontal branch the Comptonized component dominates. The disk flux scales with the inner disk temperature as T$^{5.5}$  and not as T$^{4}$ suggesting that either the inner radii changes dramatically or that the disk is irradiated by the thermal component changing its hardness factor. The PDS reveal a QPO whose frequency changes from $\sim$ 30 Hz to 50 Hz and which is found to correlate well with the disk flux ratio. In the 3-20 keV LAXPC band the r.m.s of the QPO increases with energy (r.m.s / E0.8), while the harder X-ray seems to lag the soft ones with a time-delay of a milliseconds. The results suggest that the spectral properties of the source are characterized by the disk flux ratio and that the QPO has its origin in the corona producing the thermal Comptonized component.   

      The results inferred, mainly from the LAXPC data, on various timing and spectral properties of LMXBs summarized here, demonstrate the capability of the LAXPC instrument for probing the bright as well as faint sources It is hoped that in the coming years many more LMXBs will be studied and new results will emerge from these studies.

\subsection{Studies of New Ultraluminous X-ray (ULX) Pulsars with Astrosat: }

The Ultraluminous X-ray (ULXs) sources detected in the Galaxy and other galaxies were believed for a long time to be Intermediate Mass Black Holes accreting matter from the companion star. This changed when Pulsations with 1.37 sec period were discovered from the ULX in M82 . Since then more ULX Pulsars have been discovered and at present 8 ULX Pulsars have been discovered and their properties are summarized in Table 1 taken from \citet{2020MNRAS.495.2664C}.
Astrosat discovered ULX Pulsar nature of a transient X-ray source RX J0209.6-7427 located in the Magellan Bridge when it became active after a gap of 26 years in 2019 and was studied by \citet{2020MNRAS.495.2664C} with the SXT and LAXPC instruments on Astrosat.
The transient ﬁrst detected by ROSAT during its 1993 outburst, went into a deep hibernation for 26 years and suddenly sprang back to life in 2019 with a giant outburst in 2019.  

\begin{table*}
\caption{\normalsize{Summary of the 8 known ultraluminous X-ray pulsars.}} 
\label{t3}
\centering 
\begin{tabular}{c c c c c c } 
\hline\hline 
\footnotesize{
 Name of ULX} & \normalsize{Host Galaxy} & \normalsize{Spin period (s)} & \normalsize{Orbital period (days)} & \normalsize{Spin-up/down} & \normalsize{$L_X ~(10^{39}$\,ergs\,s$^{-1})$} \\  

\hline 

 \normalsize{M82 X-2} &  \normalsize{M82} & \normalsize{1.37} & \normalsize{$\sim 2.5$} & \normalsize{Spin-up} &  \normalsize{4.9} \\  
 \normalsize{NGC 7793 P13} &  \normalsize{NGC 7793} & \normalsize{$\sim 0.42$} & \normalsize{64} & \normalsize{Spin-up} &  \normalsize{$\sim 10$} \\  
 \normalsize{NGC 5907 ULX} &  \normalsize{NGC 5907} & \normalsize{$\sim 1.13$} & \normalsize{5.3} & \normalsize{Spin-up} &  \normalsize{$\sim 100$}   \\  
 \normalsize{NGC 300 ULX1} &  \normalsize{NGC 300} & \normalsize{$\sim 31.6$} & \normalsize{-} & \normalsize{Spin-up} &  \normalsize{4.7} \\  
 \normalsize{\textit{Swift} J0243.6+6124} &  \normalsize{Milky way} & \normalsize{$\sim 9.86$} & \normalsize{$\sim 27.6$} & \normalsize{Spin-up} &  \normalsize{$\sim 2$}\\  
\normalsize{M51 ULX-7} &  \normalsize{M51} & \normalsize{$\sim 2.8$} & \normalsize{$\sim 2$} & \normalsize{Spin-up} &  \normalsize{$\sim 10$} \\  
 \normalsize{NGC 1313 X-2} &  \normalsize{NGC 1313} & \normalsize{$\sim 1.5$} & \normalsize{-} & \normalsize{Spin-up} &  \normalsize{$\sim 20$} \\  
 \normalsize{RX J0209.6-7427} &  \normalsize{SMC} & \normalsize{$\sim 9$} & \normalsize{-} & \normalsize{Spin-up} &  \normalsize{$\sim 1.6$} \\  
\hline 
\end{tabular}
\label{table:nonlin} 
\end{table*}

\begin{figure}[t!]
\begin{center}
{\includegraphics[width=1.0 \linewidth,angle=0]{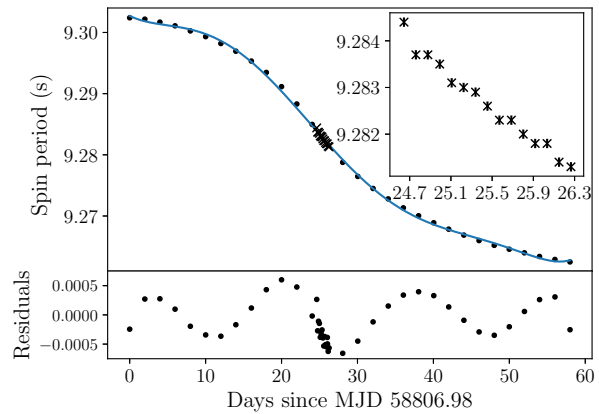}}
\end{center}
\caption{Evolution of the Pulsar period during the outburst of RX J0209.6-7427 in 2019 is shown in this plot. Spin-up of the pulsar during the outburst in clearly seen.}
\label{fig7}
\end{figure}

  Using  the SXT and LAXPC observations, \citet{2020MNRAS.495.2664C} detected strong pulsations in RX  J0209.6-7427 with 9.29 s periodicity  over a broad energy band covering 0.3-80 keV (first reported  by the NICER mission in the 0.2-10 keV energy band). The pulsar exhibited a rapid spin-up during the outburst as can be seen from Fig \ref{fig7}. Energy resolved folded pulse profiles were generated in several energy bands in 3-80 keV. This is the first report of the timing and spectral characteristics of this Be binary pulsar in hard X-rays. There is suggestion of evolution of the pulse profile with energy. The energy spectrum of the pulsar is determined and from the best-fitting spectral values, the X-ray luminosity of RX J0209.6-7427 is inferred to be 1.6$\times$10$^{39}$ erg s$^{-1}$. These timing and spectral studies suggest that this source has features of an ultraluminous X-ray pulsar in the Magellan Bridge.

\begin{figure}[b!]
\begin{center}
{\includegraphics[width=1.0 \linewidth,angle=0]{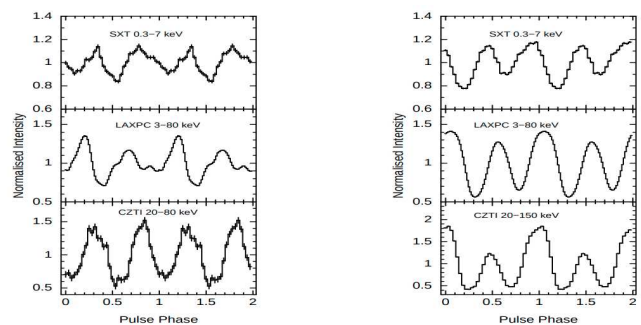}}
\end{center}
\caption{The pulse profiles derived from SXT (0.5-7.0 keV), LAXPC (3-80 keV) and CZTI (30-150 keV) are shown in the figure for the ULX  Pulsar Swift J0243.6+6124.}
\label{fig8}
\end{figure}

A second ULX Pulsar Swift J0243.6+6124, the first Galactic ultra-luminous X-ray pulsar, was observed during its 2017-2018 outburst with AstroSat at both sub- and super-Eddington levels of accretion with X-ray luminosities of L$_{x}$ $\sim$ 7$\times$10$^{37}$  and 6$\times$10$^{38}$ erg s$^{-1}$. A detailed Broadband timing and spectral  study by \citet{2020MNRAS.tmp.3039B} show that X-ray pulsations at $\sim$ 9.85 s have been detected up to 150 keV when the source was accreting at the super-Eddington level. The background subtracted pulse profiles for the  the SXT, LAXPC and CZTI instruments are shown in Fig \ref{fig8} :

The pulse profiles are a strong function of both energy and source luminosity, showing a double-peaked profile with pulse fraction increasing from 10\% at 1.65 keV to 40-80 \% at 70 keV. The continuum X-ray spectra are well-modeled with a high energy cut-off power law ($\alpha$ $\sim$ 0.6-0.7) and one or two blackbody components with temperatures of $\sim$ 0.35 keV and 1.2 keV, depending on the accretion level. No iron line emission is observed at sub-Eddington level, while a broad emission feature at around 6.9 keV is observed at the super-Eddington level, along with a blackbody radius (121-142 km) that indicates the presence of optically thick outflows.

Results on Neutron Star Binaries summarized above establish firmly the capability of the LAXPC instrument for high resolution timing and spectral studies. Large effective area enables studies of relatively fainter sources like ULX Pulsars.

\section{ Study of Black Hole X-ray Binaries (BHXBs) with LAXPC/Astrosat }

As has been mentioned earlier, the  advantages that 
 AstroSat/LAXPC has 
over RXTE are (i) an higher effective area at energies $>$ 30 keV, (ii) event
mode data allowing for temporal analysis in arbitrary user defined energy
bins and (iii) simultaneous observations from other instruments on board.
These have primarily defined and driven the LAXPC study of Black hole binaries.

It has been known that BHXBs are variable on different time-scales and
exhibit both broad band continuum noise and narrow features termed as
Quasi-periodic oscillations. While there have been several models to
explain the dynamic origin of these variabilties, a consensus and
conclusive explanation for them remains illusive. One of the possible
reasons for this impasse is that the radiative processes that are involved
in these variations have not been clearly identified. The time-averaged
spectra of black hole binaries are usually represented by physically
motivated radiative components such as disk emission, thermal comptonization,
and refection components. 

The broad band spectrum from AstroSat allows for
fitting of complex models such as General Relativistic disk emission and
blurred reflection components, which can be used to constrain physical parameters
such as the spin of the black hole and the accretion rate. It is also imperative to identify which of these
components are responsible for the variability and in particular to
identify the spectral parameters which may be changing that cause the
observed variability. It is likely that more than one of these spectral
parameters are involved and estimating any time difference between the
variations of these parameters can provide critical information regarding
the illusive dynamic origin of phenomena. AstroSat observations of BHXBs can
provide these estimates paving the way for an emerging field where variability
is quantified and understood in terms of the active radiative processes of the
system. Here we present some of the scientific results obtained so far. It should be emphasised
that these results were possible due to the unique capabilities of LAXPC  (wide band spectroscopy, higher efficiency in the hard X-rays and fine time resolution)
and the  simultaneous spectra available from SXT.

After decommissioning of {\it RXTE} in 2012, {\it AstroSat}/LAXPC is the only X-ray timing instrument best suited for studying spectral-temporal characteristics of black hole X-ray binaries (BHXBs). During last five years, LAXPC successfully  observed  many BHXBs and  very interesting results have come out. Here we discuss some of  BHXBs  results which were observed on several occasions with LAXPC instrument and in-depth analysis were performed.
In \S 3.1-3.5 we present the primary results from LAXPC grouped on the basis
of broad topics which highlight the advantage of the instrument. In \S 3.6,
we enumerate these results in a different degree of detail,
for the individual black hole sources that have been observed by LAXPC.

\subsection{Energy dependent temporal properties:}

Typically the variability is quantified by the power spectrum which is
the square of the Fourier transform of a lightcurve obtained over a certain
energy range. If normalized in a certain way, the integration of the
power spectrum (over some frequency range) gives the variance
of the variation in that energy and frequency range. Another useful quantity is
the fractional root mean square, (frms) which is the square root of the variance
normalized the mean of the lightcurve. It is insightful to consider this
as a Data Cube where the power is provided as a function of frequency and
energy. Collapsing the Data Cube over energy (or a given range of energy)
provides the power spectrum in that energy range. On the other hand
collapsing the Data Cube over a given range frequency, provides the
frms as a function of energy for that frequency range. Moreover,
cross-frequency analysis between energy bands provides time-lag as
a function of energy.

LAXPC can efficiently provide frms and time-lag as a function of energy
for different frequency bands as shown in the early analysis of the
black hole systems GRS 1915+105 \citep{2016ApJ...833...27Y} and Cygnus X-1 \citep{2017ApJ...835..195M}
Encouraged by these unprecedented results, attempts were made to develop
formalisms which could obtain physically meaningful radiative parameter
variations and time-lag between them. In the hard state, the spectra
of Cygnus X-1 above 3 keV, is dominated by thermal Comptonization.
\citet{2019MNRAS.486.2964M}. developed a model to predict the variability of such a
spectrum when the seed photon flux and the heating rate of the
corona varies with a time-lag between them.  Fitting the predictions
to the observed time-lag and  frms as a function of energy, led them
to quantify the variability in the seed photon flux and the coronal
heating rate as a function of frequency. \citet{2020ApJ...889L..17M} used the
same analysis to constrain the variability of these quantities
for the bright transient X-ray binary,  MAXI J1820+070 in its hard state
observation by AstroSat/LAXPC allowing for a comparison with Cygnus X-1.
While these analysis are for the continuum variability, \citet{2019ApJ...887..101J}
applied the model to a QPO observed by AstroSat LAXPC in the hard state
of the black hole system, SWIFT J1658.2-4242.

The model used in these works mentioned above,
is only applicable to the thermal Comptonization component, while a
thermal component is often observed in the spectra of black hole systems
which are not in a pure hard state. A generalized method which can predict
energy dependent variability properties for arbitrary spectral components is
challenging since (i) the response of the spectrum to parameter variation
has to be done numerically and (ii) the spectral parameters have to
be cast into physical ones rather than empirical ones (for e.g. heating
rate of corona instead of say the asymptotic power-law index for Comptonization
models). An initial step in this direction has been taken by \citet{2020MNRAS.498.2757G}
who have fitted the energy dependent frms and time-lag of the QPO observed
in GRS 1915+105 by AstroSat/LAXPC.

\subsection{Variations of timing and spectral properties and their correlation : }
An important and promising use of AstroSat/LAXPC data is the study
of the variation of timing features and their correlation with spectral
parameters. The sensitivity of LAXPC to detect small variation in rapid timing
properties was proved by the first time detection of the small $\sim 7$\% variation
of the high frequency ($\sim 70$ Hz) QPO in GRS 1915+105 \citep{2019MNRAS.489.1037B}, which although small
has the potential to differentiate between generic models. The QPO has also been reported
by \citet{2019MNRAS.487..928S} who showed that its presence depends on the strength of
the high energy spectral component.

The correlation with spectral
parameters was demonstrated by \citet{2019MNRAS.488..720B} where a long monitoring of
the black hole transient, MAXI J1535-571, showed a tight correlation between the
QPO frequency with the high energy spectral index rather than the flux.  An important expected correlation is that between QPO frequency and the inner radius of a truncated disk, since a physical characteristic time-scale associated with the radius maybe responsible for the QPO.
 Using LAXPC and SXT observations of the black hole system GRS 1915+105, \citet{2020ApJ...889L..36M} showed that there is such a correlation and identified the QPO frequency with the dynamical time-scale corrected by General Relativity as predicted decades ago. This important result was possible because of the broad band spectral coverage by SXT and LAXPC and the simultaneous rapid timing information provided by LAXPC.

\subsection{Long term variability :}

Long observations of BHXBs by LAXPC, which could be continuous exposure for more than a day, or monitoring of sources on weeks/months time-scale have provided extensive information regarding the long term behaviour of these sources. A good example is the enigmatic black hole system GRS 1915+105 and AstroSat was fortunate to observe it during  a transition from a non-variable class to a structured large amplitude one \citep{2019ApJ...870....4R}. This allowed for tracking of the QPO  and its energy dependent property rms and time-lag as the source made the transition. The transient Swift J1658.2-4242 shows 'flip-flop' state transitions on time-scale of minutes, which is reflected both in the spectra and power density spectrum was studied by LAXPC and other instruments  \citep{2020A&A...641A.101B}.

 Monitoring of an outburst from a black hole binary can be used to track the spectral  and timing evolution as has been done for two outbursts of 4U 1630-472 \citep{2020MNRAS.497.1197B}. The study inferred the appearance of the standard disk after a few hours of the burst and its persistence as the source evolved to the soft state. Monitoring observations of Cygnus X-3 by LAXPC has provided important clues on the formation of the radio jet and its connection to the accretion disk  \citep{2018ApJ...853L..11P}. The observations of Cyg X-3 provides a measurement of the orbital period and the discovery of  low frequency mHz QPO whose energy dependent rms and time-lag could be quantified \citep{2017ApJ...849...16P}.

\subsection{Broadband Spectral Fitting :}

Although LAXPC spectral resolution is lower than that of other instruments such as
Nustar, however it does provide a wide energy band especially when combined with SXT data.
Hence AstroSat observations have been used to fit complex spectral models to
constrain system parameters. For example, after fitting standard models
to the AstroSat spectra of the black hole transient  MAXI J1535-571, \citet{2019MNRAS.487..928S},
fitted the two component accretion flow model to the data and found that the
accretion rate to be nearly at the Eddington value and could further constrain the
mass of the black to be around $\sim 6 $ solar masses. For the same system, \citet{2019MNRAS.487.4221S}, used
the relativistic disk and blurred reflection model to constrain the spin, distance to
the source and mass of the black hole to be $\sim 10$ solar masses.  The different mass
estimate are due to differences in the model and other assumptions, but what is perhaps
important is the ability of these different models to make quantitative estimates
given AstroSat data. Black hole mass has also been estimated using two component
flow for the black hole binary 4U 1630-472, where the utility of long term monitoring
of a source as has been demonstrated. The extra galactic black hole systems LMC X-1
has a well  constrained distance and black hole
mass and hence AstroSat data has proved useful to constrain the black hole spin \citep{2020MNRAS.498.4404M}. Relativistic disk fitting of well studied sources like GRS 1915+105, can
be used to confirm the black hole mass and to constrain the spin \citep{sreehari2020astrosat}.

\subsection{Synergy with other observatories: }

Perhaps the most promising use of AstroSat data is when they are analyzed in
conjunction with data from other instruments of observatories. The complementary
nature of the different instruments provides a multi-faceted and unprecedented view
of the systems. Since black hole
binaries are variable simultaneous (or quasi-simultaneous) data would be optimal.

A good example of using the different capabilities of instruments was the
study of the black hole system 4U 1630-47 using the high energy grating spectra
from Chandra along with AstroSat's LAXPC and SXT data \citep{2018ApJ...867...86P}.
While the latter provided evidence for a highly ionized wind, relativistic disk model
fit to the continuum  using AstroSat data showed that the black hole is highly spinning,
thus making an interesting connection between wind outflow and black hole spin.
Another example is when the combination of high resolution Nustar data (3-80 keV) with
AstroSat (SXT, LAXPC and CZTI) (0.7-200 keV) of MAXI J1820+070  allowed for the use of more complex
and sophisticated models as compared to only AstroSat data  \citep{2020ApJ...889L..17M} or
just Nustar data.

Even when the observations are not simultaneous,
the use of multiple instruments has proved extremely worthwhile
as in the spectral and timing study of the flip-flop state transitions of Swift J1658.2-4242
using a host of instruments, XMM-Newton, NuSTAR, Swift, Insight-HXMT, INTEGRAL,  ATCA
and AstroSat \citep{2020A&A...641A.101B}.

It is fortunate that at present along with AstroSat,
there are two other observatories capable of
high time resolution operating namely NICER and INSIGHT-HXMT. The three observatories
together can provide unprecedented timing information in an energy range
(0.2-200 keV) which is about three order of magnitude
The improvement in our understanding that would be obtained by such analysis
is illustrated by \citet{2019JHEAp..24...30X}  where they performed  timing analysis of
Swift J1658.2-4242  outburst in 2018 with Insight-HXMT, NICER and AstroSat.
They found a range of  QPO activities detected by the different instruments and
quantified their dependence. The potential of such analysis is clear, especially
when one notes that the observations used by them were often not  strictly
simultaneous. It is expected that soon, joint analysis of AstroSat data with
one or both of these instruments, will be providing deeper insights into the
nature of Black hole systems. The importance having coordinated observations with all three
observatories cannot be over emphasised.

\subsection{Important results of some of BHXBs with LAXPC instrument:}
\subsubsection{GRS 1915+105 :}

GRS 1915+105 is a highly variable BHXBs \citep{yadav1999different, belloni2000model} with frequent radio  emission (often referred as Microquasar) \citep{fender2004towards, yadav2006connection}. We have monitored this source regularly. \citet{2016ApJ...833...27Y} have studied spectra and timing properties of GRS 1915+105  when the source had the characteristics of being in Radio-quiet during March 5-7, 2016.The  energy dependent power spectra reveal a
   strong low frequency (2 -- 7 Hz) Quasi-periodic oscillation (LFQPO)
   and its harmonic along with broad band noise. The QPO frequency
   changes rapidly with flux.   At the QPO
   frequencies, the time-lag as a function of energy has a
   non-monotonic behavior such that the lags decrease with energy
   till about 15--20 keV and then increase for higher energies.

   \begin{figure}[t]
\includegraphics[angle=-90,width=.8\columnwidth]{divya18.eps}
\caption{Color-color diagram during  fast transition among three X-ray states.}\label{grs1}
\end{figure}

  \begin{figure}[b]
\includegraphics[angle=-90,width=.8\columnwidth]{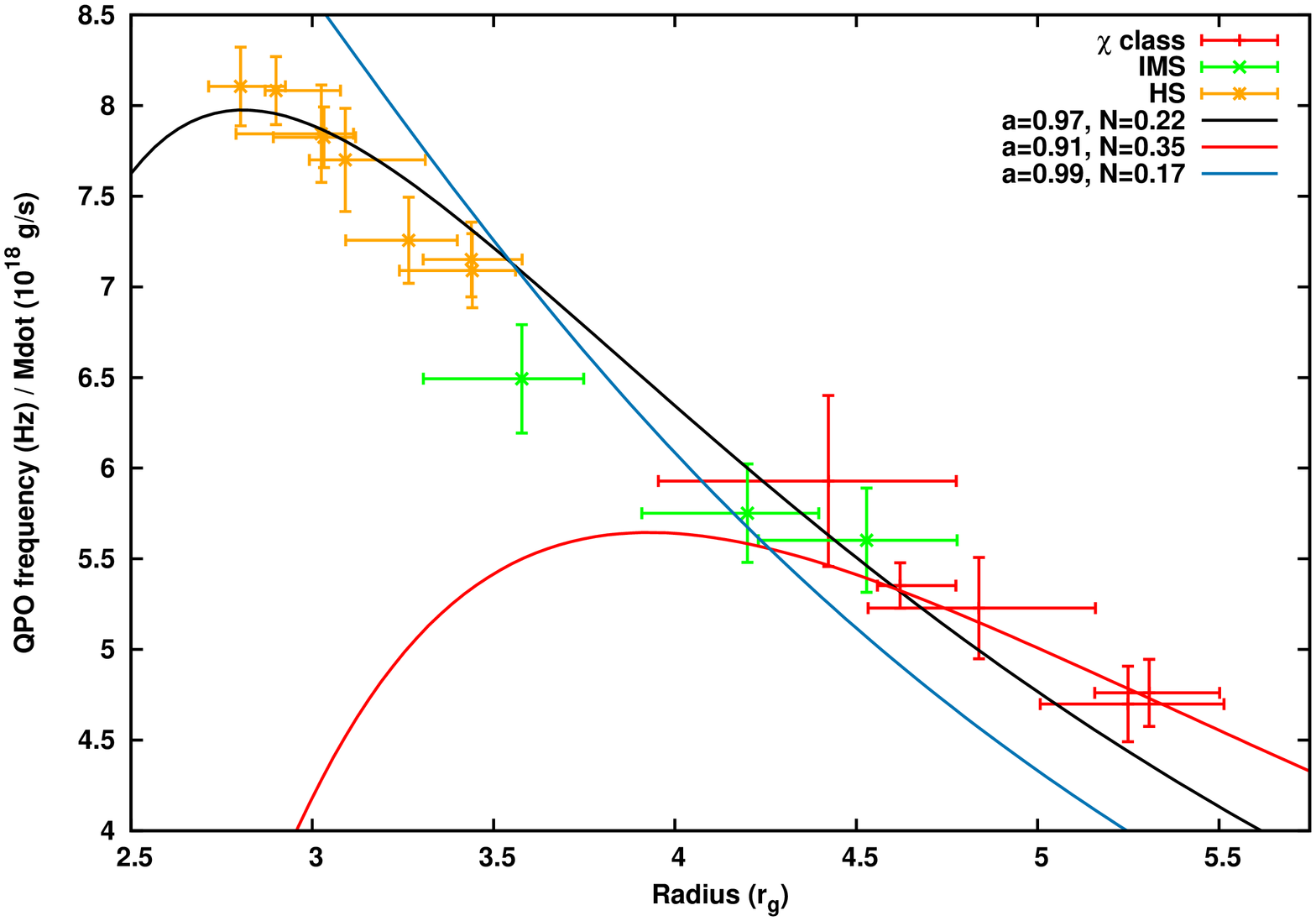}
\caption{Color-color diagram during  fast transition among three X-ray states.}\label{grs2}
\end{figure}

   \citet{2019ApJ...870....4R} have studied   GRS 1915+105 during 
   February and March, 2017  when the source underwent a major transition from a non-variable state (close to  $\chi$ class) to a  periodic flaring state (similar to $\rho$ class). It is shown that such transition takes place via an intermediate state when the large-amplitude, irregular variability of the order of thousands of seconds  turned into a 100-150 sec  nearly periodic flares similar to $\rho$ class/heartbeat oscillations. Figure \ref{grs1}  shows the color-color diagram during this fast transition. It is interesting to note that HR2 remains same when source transits to the intermediate state but when source finally attains   the flaring state, it regains the  same HR1 but HR2 has changed.  \citet{2019MNRAS.489.1037B} have  analysed  92 ks of data obtained with the LAXPC instrument  and  they  have detected around  seven percent  variation in High Frequency  QPOs in GRS 1915+105.  \citet{2020arXiv200705273B} have studied   temporal and spectral properties  of GRS 1915+105 during $\theta$  class  \citep{belloni2000model}. \citet{sreehari2020astrosat} have studied this source during the soft X-ray state and have detected HFQPs in the range  of 67.96 −- 70.62 Hz with  rms  $\sim$ 0.83 −- 1.90 percent.
   
    \citet{2020ApJ...889L..36M} showed that there is  a correlation between QPO frequency and  disk radii, and identified the QPO frequency with the dynamical time-scale corrected by General Relativity as predicted decades ago, using LAXPC and SXT  simultaneous data of  GRS 1915+105. Results of their analysis is shown in Figure \ref{grs2} which puts a tight limit on the spin of GRS 1915+105.

\subsubsection{Cygnus X-3 and relativistic large radio jets  :}     

\begin{figure}[!t]
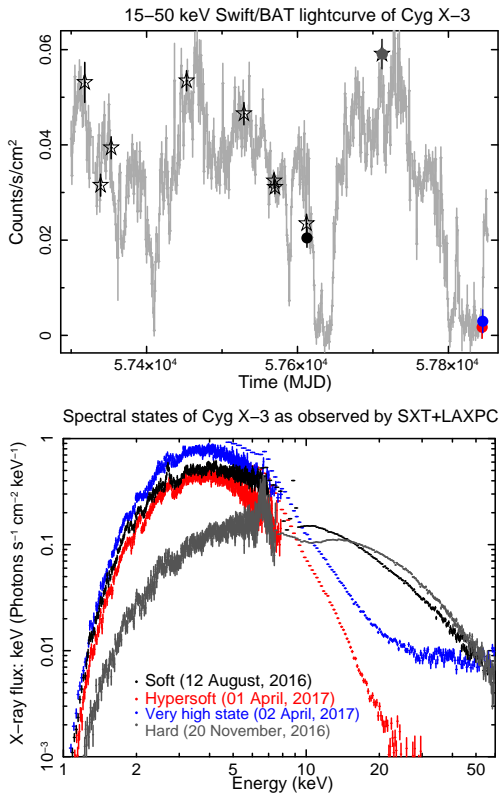

\includegraphics[angle=-90,width=.8\columnwidth]{cygx3m1.eps}
\includegraphics[angle=-90,width=.8\columnwidth]{cygx3m2.eps}
\caption{Top panel shows a typical 15-50 keV {\it Swift}/BAT lightcurve of Cygnus X-3 over a timescale of 1.5 year. {\it AstroSat}/LAXPC observations during this period are shown by stars and filled circles. Bottom panel shows unfolded spectra using SXT+LAXPC joint fitting in the energy range 0.5-60 keV. Data sets used for spectral fiting are shown with filled circle in respective colours.}\label{cygx3}
\end{figure}

Cygnus X-3 is an extraordinary HMXBs (a close binary with $\sim$ 4.8 hrs orbital period) which   produces  brightest radio jets in our Galaxy. Although the mass of the central object has not been confirmed using dynamical measurements, the similarity of its spectro-temporal properties with known black hole systems like GRS 1915+105 and XTE J1550-564 favours the binary harbouring a black hole. The source is bright and persistent in X-rays, and the X-ray emission originates as a result of wind accretion from a Wolf-Rayet companion star onto the central compact object. In contrast to the canonical BHXBs, six different spectral states have been reported in Cygnus X-3 by \citet{2010MNRAS.406..307K} using simultaneous X-ray and radio observations. Spectral states of Cygnus X-3, as observed by LAXPC are shown in the bottom panel of Figure \ref{cygx3} where unfolded, best-fit spectra from SXT and LAXPC joint fitting are shown. Details of spectral modelling are discussed in \citet{2010MNRAS.406..307K}. With LAXPC, we observed four different spectral states: soft, hard, hypersoft and very high state. The presence of the very high state is anticipated before but observed for the first time with LAXPC. Dramatic changes in flux in both soft and hard bands in different states are visible.
Very high energetic photons, of the order of GeV, have been detected during the hypersoft state when the radio emission is entirely quenched. 

{\it Astrosat}/LAXPC  performed over 20 observations of Cygnus X-3 during last five years. The top panel of the Figure \ref{cygx3} shows 15-50 keV {\it swift}/BAT lightcurve over 1.5 years and LAXPC pointing observations during this period are shown by stars and filled circles. Using three long observations during 2015-2016 spanning over several orbits, \citet{2017ApJ...849...16P} determined the binary orbital period of 17253.56 $\pm$ 0.19 sec, which is consistent with earlier measurements. However, using one-year-long observation, a slow, low-amplitude variability was observed with a periodicity of $\sim$35.8 days which may be due to the orbital precession. 
During the peak of binary orbital phase, weak but significant quasi-periodic oscillations have been observed at $\sim$5-8 mHz, $\sim$12-14 mHz and $\sim$18-24 mHz. Such detection is significant since no QPO was detected with RXTE/PCA despite its most extensive archival data. Interestingly, 7-15 mHz QPOs from Cygnus X-3 was reported with {\it Exosat}/ME observations \citep{1985Natur.313..768V}. Detailed analysis with LAXPC observations showed that QPOs were observed only during the flaring hard X-ray state and when the source is brightest (the peak phase of the binary orbital period).  It is explained in terms of increased mass accretion rate when the compact object passes through the denser wind section. An enhanced supply of material may temporarily boost the temporal variability features. 

Until now, investigating the connection between accretion disk and  the radio jets in Cygnus X-3 has been partially successful because of the lack of truly simultaneous X-ray and radio data.  X-ray and radio monitoring program of Cygnus X-3 with LAXPC provided a rare opportunity to explore the radio/X-ray connection in this  X-ray binary.

Using long-term X-ray/radio monitoring campaign with {\it Swift}/BAT and 11.2 GHz RATAN-600 telescope, it has been observed that Cygnus X-3 used to move into the hard X-ray quenched state where hard X-ray flux decreased by order of magnitude, and subsequently it showed major radio flare ejection event when the radio flux density increases dramatically from few tens of mJy up to 20 Jy. Such conjunction was detected with LAXPC and RATAN-600 telescopes on 1-2 April 2017. Using detailed X-ray analysis, \citet{2018ApJ...853L..11P} found that Cygnus X-3 undergo spectral state transition from the hypersoft  state (HPS) to a harder, more luminous state which was never observed before. We term it as the very high state (VHS). Such a transition occurred within a few hours when the radio flux density increases from $\sim$100 mJy to $\sim$478 mJy. Using SXT+LAXPC joint spectral analysis, they observed no hard X-rays above 17 keV during the HPS state. Within hours timescale, a flat power law appeared in the spectra with the power-law index of 1.49$^{+0.04}_{-0.03}$ and extended up to 70 keV. Such an observation provided direct evidence of synchrotron emission that originates in the radio-emitting blob which was caught in the act of decoupling from the accretion disk. 
Such a detailed radio/X-ray coupling event was observed for the first time and made possible due to the LAXPC's capabilities of higher efficiency for hard X-ray above 30 keV and high time resolution. 

\subsubsection{Cygnus X-1   :}

   \begin{figure}[b]
\includegraphics[angle=-90,width=.8\columnwidth]{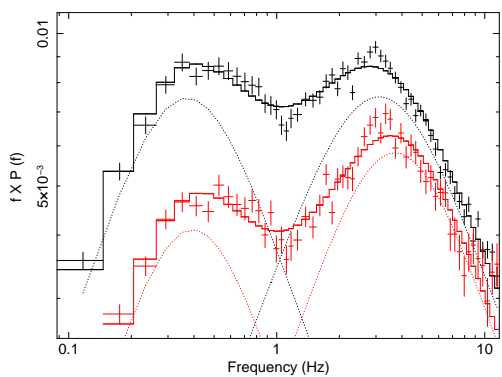}
\caption{The frequency times the power spectra of Cygnus X-1
for   two energy bands; 3-10 keV band (black points) and  20-40 keV band (red points).}\label{cyg11}
\end{figure}

 Probably one of the best studied black hole systems, the bright black hole system, Cygnus X-1, presents an excellent example to underline  LAXPC capabilities and the new understanding that LAXPC data can bring. The higher effective area at high energies has allowed LAXPC to study in detail the energy dependent rapid variability of Cygnus X-1 in the hard state. LAXPC data analysis by \citet{2017ApJ...835..195M}, revealed the rms and time-lag variation as a function of Fourier frequencies in the broad energy range of 4 -- 80 keV, extending the earlier results which were limited to 30 keV. They also showed that the event mode data of LAXPC allows for flux resolved spectroscopy in rapid ($\sim 1$) second time-scales, which for Cygnus X-1 revealed a correlation between photon index and flux. A more extensive analysis of six observations of Cygnus X-1 by \citet{2019MNRAS.486.2964M}, showed that the energy and frequency dependent rapid variability is different for observations which are in the hard state. More importantly, they could now quantitatively fit and explain this variability in terms of a single zone stochastic fluctuation model.

\subsubsection{MAXI J1535-571    :}

 The bright outburst of this recently discovered black hole binary has been observed by a large number of observatories
including AstroSat and has provided one of the most comprehensive view of any such systems. LAXPC detected prominent QPOs at $\sim 2$ Hz in a wide energy range of 3-50 keV \citep{2019MNRAS.487..928S} when the source was in the hard intermediate state. Spectral modelling using standard
Comptonization models and physically motivated ones like the two-component accretion flow revealed that source was nearly  at Eddington limit and its black hole mass is $\sim 6 M_\odot$. A detailed study of the broad band spectrum from SXT and LAXPC using relativistic reflection models provided an estimate of the black hole spin parameter ($a \sim 0.7$) and a mass of $\sim 10 M_\odot$  \citep{2019MNRAS.487.4221S}. Correlation between the QPO frequency and spectral parameters was studied by \citet{2019MNRAS.488..720B} who found that the frequency tightly correlates with the photon index rather than the flux.

\subsubsection{4U 1630-472   :}

Among recurrent black hole X-ray transients, 4U 1630$-$472 is unique as it shows frequent outbursts at about 600 days interval.  The source was also at the centre of focus due to its `superoutburst' which typically lasts for two years. Because no dynamical mass measurements have been performed, the source has been categorised as the black hole candidate due to its similarity of X-ray spectra-timing characteristics with other confirmed black hole X-ray binaries. A dust scattering halo analysis placed the source between 4.7-11 kpc \citet{2018ApJ...859...88K} while the spectral index vs mass accretion rate correlation was used to estimate the black hole mass of 10 $\pm$ 0.1 M$\odot$ \citep{2014ApJ...789...57S}.

 Two major monitoring campaigns with AstroSat/LAXPC took place between 27 August $-$ 2 October 2016 and 4 August $-$ 17 September 2018 to monitor `mini' outbursts from 4U 1630$-$472 which typically lasts 5-6 months. Using both outburst observations, \citet{2020MNRAS.497.1197B} performed detailed spectro-timing analysis. Using broadband (0.7-20.0 keV) spectral modelling, they detected no disk component during the onset of the 2016 outburst while later a geometrically thin accretion disc with the temperature of $\sim$1.3 keV was observed. Such behaviour was also accompanied by other observations like the steepening of the powerlaw index from 1.8 to 2.6 and the decrease of rms variability by $\sim$5\%.
However, all observations with LAXPC during 2018 outburst showed thermal disk emission dominated state with the disc temperature vary between 1.26 $\pm$ 0.01 keV and 1.38 $\pm$ 0.01 keV. Depending upon spectral characteristics, they observed three distinct spectral states: low hard, intermediate and high soft states. Three states can be distinguished in the `C'-shaped hardness intensity diagram track where the hardness is the ratio between the counts in 6-20 keV and 2-6 keV and the intensity is 2-20 keV X-ray flux calculated from LAXPC spectral modelling. Such distinction was also apparent in the flux-rms plot where the flux was calculated from the best-fit spectra while the rms was calculated using 0.005-10 Hz power density spectra.
The transition among three states is in agreement with the accretion geometry inferred from the two-component accretion flow model (Chakrabarti and Titarchuk 1995). 

Due to its high energy coverage, LAXPC spectra of 4U 1630-472 were also found useful in measuring fundamental properties of the accreting compact object when combined with concurrent spectra from other soft X-ray instruments like SXT and Chandra. For example, the spin of a black hole  was measured using luminous high soft state observation with AstroSat and Chandra. Using continuum spectral modeling method, \citet{2018ApJ...867...86P} determined the presence of a fast-spinning black hole in 4U 1630-472. Using relativistic continuum spectral modeling on three independent measurements with Chandra, SXT+LAXPC and Chandra+SXT+LAXPC and applying Markov Chain Monte Carlo simulations on fitted spectral parameters, they constrained the spin of the black hole to be 0.92 $\pm$ 0.04 within 99.7\% confidence limit. Such a measurement is essential for explaining accretion inflow and outflow properties very close to the accreting black hole. 

\subsubsection{MAXI J1820+070  :}

At a distance of 3.46$^{+2.18}_{-1.03}$ kpc \citep{2020MNRAS.496L..22G}, MAXI J1820+070 was discovered as one of the closest and the brightest Galactic BHXBs known till date \citep{2016A&A...587A..61C}. With the Swift/BAT peak flux of $\sim$4 Crab in 15-50 keV, the source showed back and forth transition between hard and soft states. The source was proposed as a black hole candidate depending upon the large amplitude hard X-ray variability observed in the power density spectrum \citep{2018ATel11423....1U}. However, with the MAXI data, `q' shaped track is clearly visible in the HID \citep{2020MNRAS.498.5873C} strengthening the fact that the source harbours a black hole. AstroSat observed the source on two occasions during the peak of the outburst. 

Using the  two days long observation with LAXPC on 30 March 2018, \citet{2020ApJ...889L..17M} observed 47.7 mHz QPO in the power density spectra which is similar to what observed from two other BHXBs: GRS 1915+105 and IGR J17091-3624. Using an SXT+LAXPC joint spectral analysis in the energy range 0.7-30 keV, they observed a flat spectral index of $\sim$1.61 and evidence for cool disk at the temperature of $\sim$0.22 keV. They noted that the time lag (where the reference band is 4.15-5.37 keV) increasing linearly with the photon energy at 0.1, 1 and 10 Hz. However, the fractional rms decreases with the energy at 0.1 and 1 Hz and independent of the photon energy at 10 Hz. The observed temporal properties, e.g., typical 50-100 msec time lag, can be explained qualitatively by the stochastic propagation fluctuation model \citep{2019MNRAS.486.2964M}. Such a detailed temporal analysis brings out the importance of LAXPC observations.

\citet{2020MNRAS.498.5873C} studied the broadband spectra which include SXT, LAXPC and CZTI in the energy range of 0.3$-$120 keV
and the best fit spectral modelling shows the presence of a soft excess component in the form of thermal disk blackbody which is significantly below three keV, a thermal comptonisation model with a very flat spectral index close to 1.4 and two relativistic reflection models to fit broad and narrow iron line complex and a Compton hump. Two relativistic models differ significantly in their coronal electron temperature by a factor of $\sim$7. Interestingly, the relativistic reflection component dominates the broadband spectra. Comparing with the detailed NuSTAR spectral study by \citet{2019MNRAS.490.1350B}, they inferred that the corona in MAXI J1820+070 is inhomogeneous and residing close to the black hole ($<$ 3 R$_g$).

\section{Active Galactic Nuclei}

AstroSat/LAXPC has the potential to monitor the variability of Active
Galactic Nuclei (AGN) and in conjunction with SXT and UVIT to study
the their broad band spectra. Blazars are jet dominated AGN which
show high amplitude variability. Using AstroSat LAXPC and SXT
observations \citet{2020arXiv200705273B} constructed the power spectrum for the lighcurve
of the blazar, Mrk 421, and detected a break or a characteristic time-scale. Such
characteristic time-scale have been detected before for X-ray binaries and
regular AGN and are believed to originate in the accretion disk. Hence this
result suggests that the jet variability also has an disk origin.
For the blazar, RGB J0710 + 591, \citet{2020MNRAS.492..796G} reported a significant deviation from a power-law shape for the X-ray spectrum obtained from LAXPC and SXT.
Such deviation or curvature reflect the underlying shape of the particle
energy spectrum that produces the emission. 

The AstroSat results suggested
that particle distribution has a maximum energy cutoff as predicted by
models where the particles are shock accelerated and radiatively cooled.
\citet{2020MNRAS.492..796G} studied the broad band spectral energy distribution
for the blazar 4C +21.35 using multi-wavelength data from AstroSat and
other observatories. They modeled the spectrum during flaring and
quiescent states with two compact regions which originated at different
times and moved away from the central region. For the regular
AGN, RE J1034+396, \citet{2018MNRAS.478.4830C} fitted the LAXPC and SXT
spectra to reveal the presence of a soft excess, consistent with earlier
results.

Studies of faint sources like AGN by LAXPC is limited by the
accuracy of the instrument's background estimation. It should
be noted that most of the results of AGN data analysis using the RXTE/PCA
observations were undertaken in the later times of the satellite operations
when the response and background were better characterised.
With improvements and different estimation techniques for LAXPC
background \citep{2020JAPA.tmp.antia, 2020JAPA.tmp.misra}                  this issue,
it is expected that in the near future there will be significant spectral and variability studies of AGN using LAXPC data. 

\section{Summary}
      LAXPC instrument has carried on the legacy of  RXTE PCA and HEXTE  by
being one of the primary instruments to study rapid variability of X-ray
systems. As compared to RXTE/PCA, it has enhanced features which are
(i) an higher efficiency at hard X-rays (energies $\ge$ 30 keV), (ii) event mode data allowing for temporal analysis in arbitrary user defined energy bins and (iii) simultaneous SXT data at the soft X-ray band (0.5--8 keV). These advantages have
been demonstrated by a  number of publications reporting important results discussed in this paper for more than 30 sources. Moreover, the increasing reservoir of archived and future
LAXPC data  is expected to provide an unprecedented view of known sources as well as those that are yet to be discovered.
Improvements in the calibration, especially in the background estimation, would
lead to more detailed spectral and timing analysis of the bright X-ray sources and will enable analysis of fainter sources such as a larger number of
active galactic nuclei.

It is fortunate that at present there are several X-ray missions in operation,
which have complimentary capabilities to the instruments on-board AstroSat.
Several works have already demonstrated the dramatic advantage of having
simultaneous (or even quasi-simultaneous) observations of a source with multiple X-ray missions and with those in other wavebands. There is now an unprecedented opportunity of detailed timing studies for a wide energy range using simultaneous observations of AstroSat/LAXPC and NASA’s Nicer. Despite the challenges of
organizing and administrating such coordinated observations by multiple missions (Nicer, Nustar, chandra and other),
it is imperative for break through results, that a large number of them should be undertaken.

\section*{Acknowledgements}
We acknowledge the strong support from Indian Space
Research Organization (ISRO) in various aspects of instrument development, space qualification, software development,
mission operation and data dissemination. We  specially acknowledge ISAC support for electronics development and during space qualification tests. We thank IISU tean for providing us bellow pump along with space qualified pump driver.  We acknowl-
edge support of the scientific and technical staff of the LAXPC instrument team for their excellent team work as well as staff of the TIFR Workshop who helped us at various level of LAXPC instrument development.



\end{document}